\documentclass[twocolumn,showpacs]{revtex4}
\usepackage{amssymb}
\usepackage{amsmath}
\usepackage{dcolumn}
\usepackage{bm}
\usepackage{graphicx}

\begin{document}

\title{Quantum simulation of artificial Abelian gauge field using
nitrogen-vacancy center ensembles coupled to superconducting resonators}
\author{W. L. Yang$^{1,4}$}
\author{Zhang-qi Yin$^{2,3}$}
\email{yinzhangqi@gmail.com}
\author{Z. X. Chen$^{3}$}
\author{Su-Peng Kou$^{5}$}
\author{M. Feng$^{1}$}
\email{mangfeng@wipm.ac.cn}
\author{C. H. Oh$^{4}$}
\email{phyohch@nus.edu.sg}
\affiliation{$^{1}$State Key Laboratory of Magnetic Resonance and Atomic and Molecular
Physics, Wuhan Institute of Physics and Mathematics, Chinese Academy of
Sciences, and Wuhan National Laboratory for Optoelectronics, Wuhan 430071,
China}
\affiliation{$^{2}$Key Laboratory of Quantum Information, University of Science and
Technology of China, Chinese Academy of Sciences, Hefei 230026, China }
\affiliation{$^{3}$Center for Quantum Information, IIIS, Tsinghua University, Beijing,
China}
\affiliation{$^{4}$Centre for Quantum Technologies, National University of Singapore,
Singapore 117543, Singapore}
\affiliation{$^{5}$Department of Physics, Beijing Normal University, Beijing 100875, China}

\begin{abstract}
We propose a potentially practical scheme to simulate artificial Abelian
gauge field for polaritons using a hybrid quantum system consisting of
nitrogen-vacancy center ensembles (NVEs) and superconducting transmission
line resonators (TLR). In our case, the collective excitations of NVEs play
the role of bosonic particles, and our multiport device tends to circulate
polaritons in a behavior like a charged particle in an external magnetic
field. We discuss the possibility of identifying signatures of the
Hofstadter "butterfly" in the optical spectra of the resonators, and analyze
the ground state crossover for different gauge fields. Our work opens new
perspectives in quantum simulation of condensed matter and many-body physics
using hybrid spin-ensemble circuit quantum electrodynamics system. The
experimental feasibility and challenge are justified using currently
available technology.
\end{abstract}

\maketitle

Gauge field theory \cite{Jean}, which was developed for describing the
subatomic interaction, has deepened our understanding of a wide range of
physical phenomena: quantum magnetoresistance oscillations, superconducting
vortices, quantum Hall effects (QHE), and Hofstadter butterfly (HB) \cite{HB}
\textit{etc}. However, these phenomena associated with the gauge field are
very difficult to be directly observed in ordinary solid-state systems due
to the requirement of extremely high field. Recently, following the idea of
quantum simulation \cite{Q1}, several theoretical schemes have been proposed
to generate artificial gauge field in atomic, molecular, and optical
systems. The first route is taken in cold atomic systems. The gauge field
can be generated by rotating the trap or lattice \cite{RL,RB}, by
introducing appropriate phase factors for hopping amplitudes through
combining Raman-laser-assisted tunneling and lattice acceleration methods
\cite{Hall,AB,FS,Me1,Me2}, or by tailoring orthogonal laser-beam coupled to
the atomic degenerate internal states \cite{Ma,Me3}, or by simply using the
ordinary tunneling in an optical lattice \cite{Ch2011}. The second route is
based on cavity quantum electrodynamics (QED) systems, where the gauge field
can also appear through manipulating the phase factors of the hopping term,
in circuit QED cavity arrays \cite{Koch}, in confined ions in microtrap
array \cite{Ber}, in coupled resonator optical waveguides \cite{Haf}, and in
solid-state photonic structures \cite{Um}. In addition, the fractional
quantum Hall effect (FQHE) can also be simulated in optical-cavity arrays by
trapping three-level atoms and using elaborated laser driving \cite{E2}.
However, due to the required rigorous conditions, up to now only few
pioneering experiments have successfully mimicked the effect of a
light-induced artificial magnetic field in an optical lattice \cite{Aide}
and a synthetic electric field Bose-Einstein condensates \cite{Lin2} by
using a Raman-assisted tunneling method \cite{Me1}, and an effective
time-dependent vector potential, respectively.

\begin{figure}[tbp]
\includegraphics[width=7 cm]{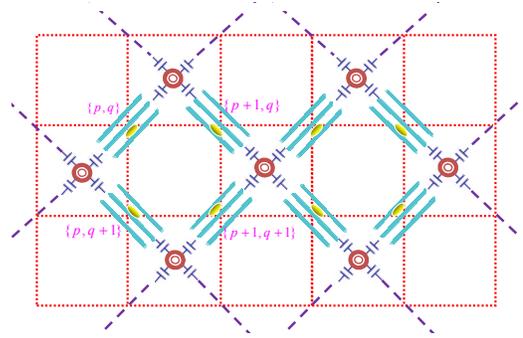}
\caption{(Color online) Schematic circuit for the resonator array, where the
NVEs are magnetically coupled to the quantized field of the TLR, and the
circles denote the central coupler. Microwave photons can tunnel between
adjacent resonators via capacitive coupling. The lattice sites are
represented by $a(p\hat{x}+q\hat{y})$ with $a$ the spacing of the lattices,
where the dotted grid denote the 'vertical' and 'horizontal' direction.}
\end{figure}

In this paper we propose an alternative theoretical approach for quantum
simulating an effective gauge potential in a hybrid solid-state system. We
consider NVEs \cite{N1,N2,N3,N4,N5,N6,N7} which are confined in
two-dimensional ($2D$) square lattices of superconducting microwave TLR \cite%
{SC}. The NVEs are driven by series of microwave sources with different
phases at different sites. We carefully tune the frequencies of the cavity
modes, the driving strength and the phase of microwave sources, in order to
induce the non-zero tunneling phases around a closed loop in real space and
to generate a non-vanishing artificial gauge field experienced by the
orbital motion of the polaritons in NVEs. Our approach utilizes a
space-dependent coupling between internal ground states of the NVEs, which
yields crucial phases to create considerable synthetic gauge fields. This
intriguing behavior is indeed the analogue of the motion of charged particle
in an real magnetic field.

The main merits of the present system include the \textit{in situ}
tunability of the parameters of the circuit elements, individual addressing,
the peculiar characteristics of the NVE \cite{NV} (e.g., long coherence time
at room-temperature), and the scalability of cavity resonator arrays \cite%
{Koch,Nu1,Nu2}. Very recently, D. Underwood \textit{et al} experimentally
demonstrated 25 arrays of twelve capacitively coupled TLRs and accessed the
feasibility of quantum simulation in cavity QED systems \cite{Und}.
%The recent experimental demonstration of an excellent
%quantum control over photonic Fock states in three resonators interconnected
%by two phase qubits \cite{Nu1} shows the extension from single to more
%versatile multi-resonator architectures, allowing manipulation of spatially
%separated bosonic modes, which demonstrates single-photon Rabi swaps between
%two resonators detuned by $\sim 12000$ resonator linewidths.
It implies that the polariton-polariton interaction between distant NVEs can
be effectively tuned in a controllable way, which renders our scheme to be
more practical. By combining the related spectroscopic circuit QED technique
for readout of the quantum states of individual constituent elements, we
could probe the properties of the system by independently detecting the
correlation between distant sites or output fields of the TLRs, which
provides the feasibility of observing the strong gauge field effects, such
as HB spectrum, in a realistic hybrid solid-state system.

\begin{figure}[tbp]
\includegraphics[width=5 cm]{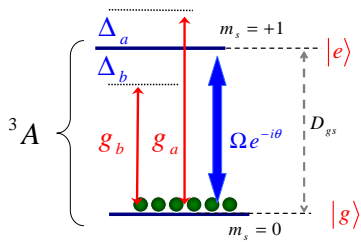}
\caption{(Color online) Level structure of a NVE, where the electronic
ground state is an electron spin triplet state (S=1), and $D_{gs}/2\protect%
\pi =2.87$ GHz is the zero-field splitting between the $m_{s}=0$ sublevel
and the $m_{s}=\pm 1$ sublevels in the absence of external magnetic field.
The degeneracy between states $m_{s}=\pm 1$ is lifted by applying a static
magnetic field with 40 $mT$ parallel to the chip and along the $[111]$
crystalline axis within a few degrees. With this orientation, the four
equivalent NV center crystalline orientations $\langle 111\rangle $ all make
approximately the same angle $55^{\circ }$ with the magnetic field so that
their resonance frequencies are approximately equal. The transition between $%
\left\vert g\right\rangle $ and $\left\vert e\right\rangle $ with Larmor
frequency $\protect\omega _{e}$ is coupled to the two-mode TLR with the
coupling rates $g_{a}$ ($g_{b}$) and detunings $\Delta _{a}$ ($\Delta _{b}$%
), and is also driven by a classical field with the Rabi frequency $\Omega $
and the related phase $\protect\theta $.}
\end{figure}

As illustrated in Fig. 1, the system we study is a $2D$ square lattice,
where the basic unit consists of a NVE confined in a microwave-driven
two-mode TLR with the length $L$, the inductance $F_{t}$ and the capacitance
$C_{t}$, a narrow center conductor and two nearby lateral ground planes.
Each NVE has the average NV center density $1\times 10^{15}$ $cm^{-3}$ \cite%
{exp}, where each NV center is negatively charged with two unpaired
electrons located at the vacancy, which can be modeled as a two-level system
in the ground-state subspaces as shown in Fig. 2. For clarity of
description, we adopt following denotations: $\left\vert
^{3}A,m_{s}=0\right\rangle =\left\vert g\right\rangle $, and $\left\vert
^{3}A,m_{s}=+1\right\rangle =\left\vert e\right\rangle $ (we distinguish the
degenerate sublevels $\left\vert ^{3}A,m_{s}=\pm 1\right\rangle $ by
appropriate external magnetic field or polarized irradiation). In our
system, microwave-photons can tunnel between adjacent TLRs via capacitive
coupling by connecting TLRs via a central coupler, which serves as
individual tunable quantum transducers to transfer photonic states between
the resonators.

The overall system is governed by the Hamiltonian $H_{tot}=%
\sum_{p,q}(H_{C}^{(p,q)}+H_{JC}^{(p,q)}+H_{NV}^{(p,q)}+H_{T}^{(p,q)})$,
where $H_{C}^{(p,q)}=\sum_{c=a,b}\hbar \omega _{c}c_{p,q}^{\dagger }c_{p,q}$
is for the $(p,q)$-th TLR's mode with $c_{p,q}^{\dagger }$ $(c_{p,q})$ the
creation (annihilation) operators of the full-wave mode of the resonator and
$\omega _{c}=2\pi /(\sqrt{F_{t}C_{t}})$ the corresponding eigenfrequency.
The collectively magnetic coupling between NVE and all the microwave modes
can be modeled as
\begin{equation}
H_{JC}^{(p,q)}=\sum_{p,q}\sum_{c=a,b}\hbar
(g_{p,q}^{c}S_{p,q}^{+}c_{p,q}+\Omega _{p,q}e^{i(\omega _{l}t+\theta
_{p,q})}S_{p,q}^{+}+H.C.),
\end{equation}%
where $g_{p,q}^{c}$ is the single NV center's vacuum Rabi frequency in the $%
(p,q)$-th resonator for mode $c$, and $\omega _{l}$ is the frequency of the
external driving laser with the Rabi frequency $\Omega _{p,q}$ and the phase
$\theta _{p,q}$. The collective transition operators of NVE are defined as $%
S_{p,q}^{+}=\sum_{j=1}^{N_{p,q}}|e\rangle _{p,q}^{j}{}_{p,q}^{j}\langle g|$
and $S_{p,q}^{-}=\sum_{j=1}^{N_{p,q}}|g\rangle _{p,q}^{j}{}_{p,q}^{j}\langle
e|$ with $N_{p,q}$ the number of the NV centers in a NVE inside the $(p,q)$%
-th resonator.

The Hamiltonian of a NVE reads $H_{NV}^{(p,q)}=$ $\frac{\hbar }{2}\omega
_{e}S_{p,q}^{z}$ with $\omega _{e}$ the energy-level spacing of the states $%
\left\vert e\right\rangle $ and $\left\vert g\right\rangle $. All the spins
in NVE interact symmetrically with a single mode of electromagnetic field by
magnetic-dipole coupling because the mode wavelength is larger than the
spatial dimension of the NVE if the spin ensemble is placed near the TLR's
field antinode. We suppose that the mode $a_{p,q}$ couples with the vertical
adjacent sites, and $b_{p,q}$ couples with the horizontal adjacent sites. So
the tunneling between adjacent sites can be governed by the Hamiltonian $%
H_{T}^{(p,q)}=-\hbar (T_{a}a_{p,q+1}^{\dagger }a_{p,q}+T_{a}a_{p,q}^{\dagger
}a_{p,q+1}+T_{b}b_{p+1,q}b_{p,q}^{\dagger }+T_{b}b_{p,q}b_{p+1,q}^{\dagger
}).$ Here the intercavity photonic tunneling rates $T_{a}(T_{b})$ are
tunable experimental parameters due to the flexibility of the central
coupler. The central coupler may be conceived as a current-biased Josephson
junction phase qubit \cite{CBJJ}, or Josephson ring circuit \cite{Koch}, or
a capacitive coupling element \cite{Hu}, or an active non-reciprocal devices
as proposed in \cite{Kam}.

Under the strong driving case $\Omega _{p,q}\gg T_{c}$, $\sqrt{N_{p,q}}
g_{p,q}^{c}$, we define the new energy levels as $|+\rangle
_{p,q}=(e^{i\theta _{p,q}}|e\rangle _{p,q}+|g\rangle _{p,q})/\sqrt{2}$ and $%
|-\rangle _{p,q}=(-e^{i\theta _{p,q}}|e\rangle _{p,q}+|g\rangle _{p,q})/
\sqrt{2}$. Introducing $\Delta _{c}=\omega _{c}-\omega _{e}$ and $\delta
_{p,q}^{c}=\Delta _{c}-\Omega _{p,q}$, and supposing $|\delta _{p,q}^{c}|\ll
\Omega _{p,q}$, we have the effective Hamiltonian of the mode $a$ in lattice
site $(p,q)$ under the new basis as $H_{p,q}^{a}=\delta
_{p,q}^{a}a_{p,q}^{\dagger }a_{p,q}+(\frac{1}{2}g_{p,q}^{a}e^{-i\theta
_{p,q}}a_{p,q}\tilde{S} _{p,q}^{+}+\mathrm{H.C.})$, with $\tilde{S}%
_{p,q}^{+}= \sum_{j=1}^{N_{p,q}}|+\rangle _{p,q}^{j}{}_{p,q}^{j}\langle -|$
(see Sec.I in supplementary material). On the other hand, the effective
Hamiltonian describing the mode $b$ is $H_{p,q}^{b}=\delta
_{p,q}^{b}b_{p,q}^{\dagger }b_{p,q}-(\frac{1 }{2}g_{p,q}^{b}e^{-i\theta
_{p,q}}b_{p,q}^{\dagger }\tilde{S}_{p,q}^{+}+ \mathrm{H.C.}).$

For simplicity, we suppose $\delta _{p,q}^{c}=\delta $, $T_{c}=T$, and $%
g_{p,q}^{c}=g$ for all $p,q,c$. In the limit of $\delta \gg T$, $g$, we can
adiabatically eliminate the photonic modes and get the effective Hamiltonian
between NVEs (see Sec.II in supplementary material),
\begin{equation}
\begin{aligned} H_{\mathrm{eff}}=&-J^{'}(\sum_{p,q}
e^{i(\theta_{p,q}-\theta_{p+1,q})}\tilde{S}_{p,q}^+ \tilde{S}_{p+1,q} \\&+
e^{i(\theta_{p,q+1}-\theta_{p,q})}\tilde{S}_{p,q}^+ \tilde{S}_{p,q+1} +
\mathrm{H.C.}) \end{aligned}
\end{equation}
where $J^{^{\prime }}=T(g/2\delta )^{2}$.

Using the Holstein-Primakoff transformation $\tilde{S}^{+}=B_{p,q}^{\dagger}%
\sqrt{N_{p,q}-B_{p,q}^{\dagger }B_{p,q}}\simeq \sqrt{N_{p,q}}%
B_{p,q}^{\dagger }$, $\tilde{S}_{p,q}^{-}=B_{p,q}\sqrt{N_{p,q}-B_{p,q}^{%
\dagger }B_{p,q}}\simeq \sqrt{N_{p,q}}B_{p,q}$, and $\tilde{S}%
_{p,q}^{z}=(B_{p,q}^{\dagger }B_{p,q}-N_{p,q}/2)$ \cite{HP}, where the
operator $B_{p,q}^{\dagger }(B_{p,q})$ fulfills the bosonic commutation
relation $[B_{p,q},B_{p^{\prime },q^{\prime }}^{\dagger }]\simeq \delta
_{pq,p^{\prime }q^{\prime }}$ in the case of low number of NVE excitations,
we can map the collective raising (lowering) operators $\tilde{S}_{p,q}^{\pm
}$ of the $(p,q)$-th NVE into the bosonic operators $B_{p,q}^{\dagger }$ and
$B_{p,q}$. These transformations change the Hamiltonian $H_{eff}$ into a
concise form as
\begin{eqnarray}
H_{B} &=&-J(\sum_{p,q}e^{i(\theta _{p,q}-\theta _{p+1,q})}B_{p,q}^{\dagger
}B_{p+1,q}  \notag \\
&&+e^{i(\theta _{p,q+1}-\theta _{p,q})}B_{p,q}^{\dagger }B_{p,q+1}+\mathrm{\
H.C.}),
\end{eqnarray}%
where $J=N_{p,q}J^{^{\prime }}=T(g_{eff}/2\delta )^{2}$ and the collective
coupling rates $g_{eff}$ between NVE and TLR in each lattice site are
supposed to be equal in our case.

\begin{figure}[tbp]
\includegraphics[width=4.25cm]{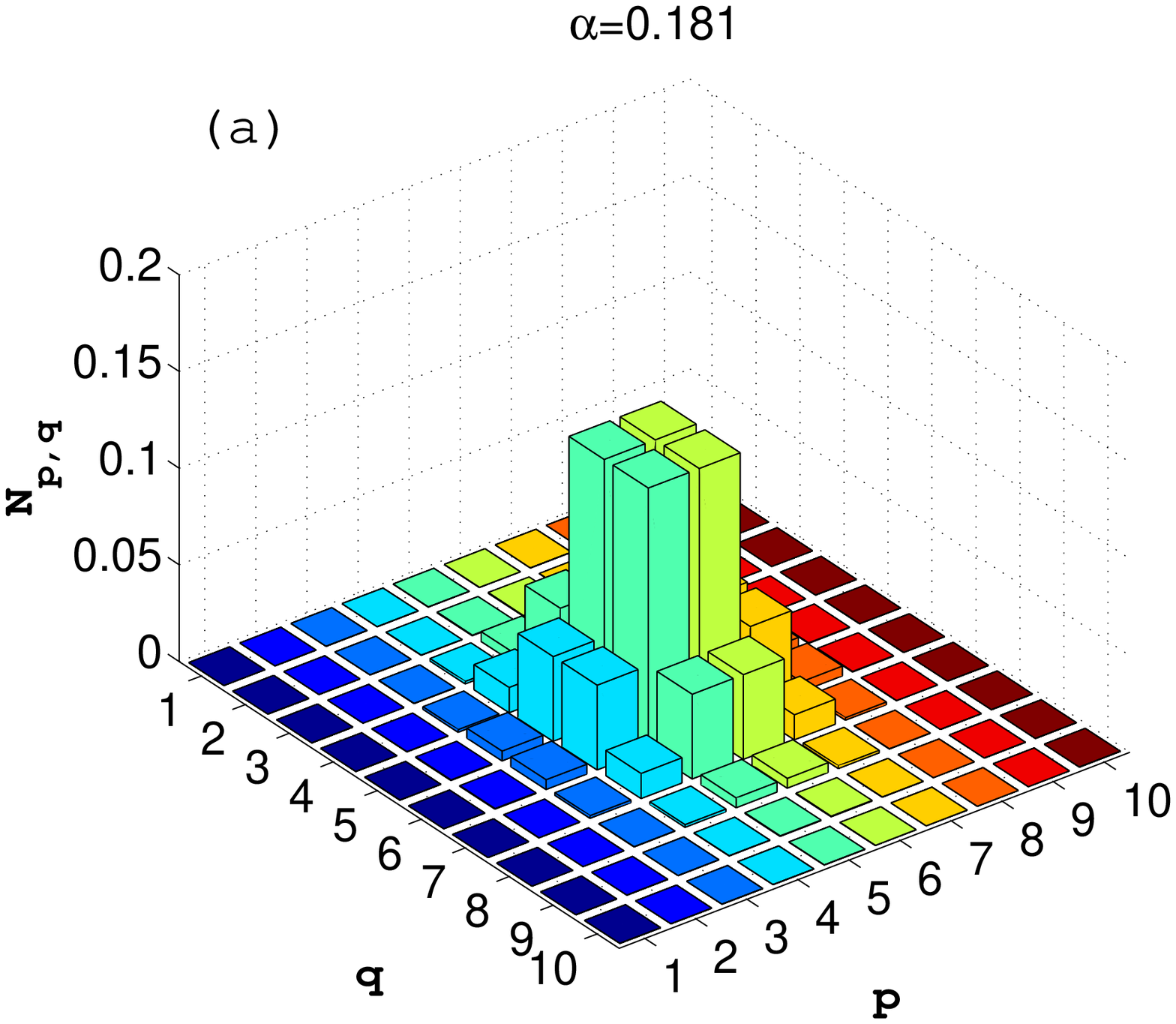} %
\includegraphics[width=4.25cm]{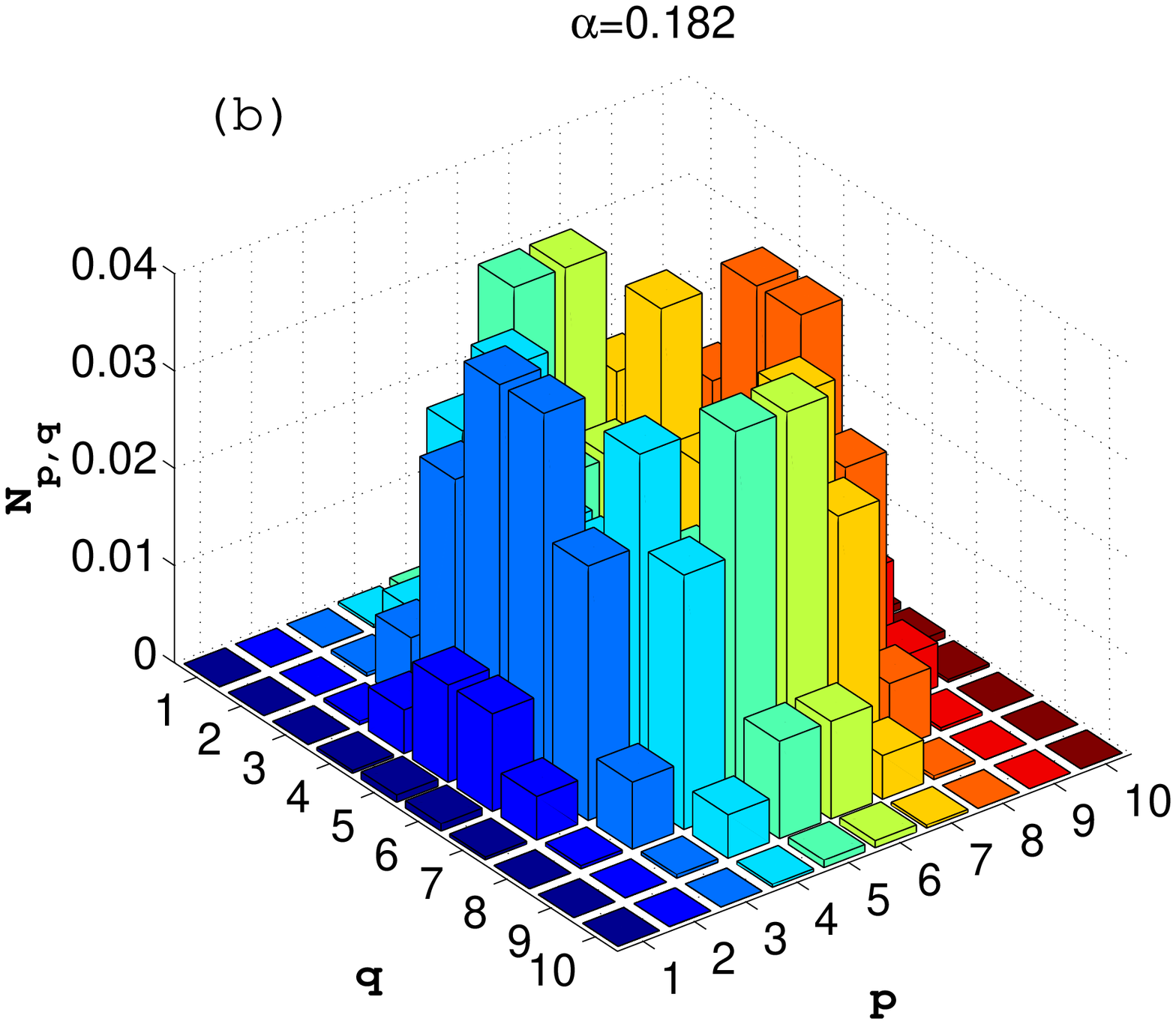}
\caption{(Color online) For $10\times 10$ lattice, the spatial distribution
of absolute square of the polariton $B_{p,q}$ wave function in the ground
state of Hamiltonian (3) with open boundary conditions, under different
values of $\protect\alpha $: (a) $\protect\alpha =0.181$. (b) $\protect%
\alpha =0.182$. }
\end{figure}

Because of the spatial variation of the tunneling phase, the wave function
of a polariton from one lattice site to another acquires a nontrivial phase,
which can be interpreted as an effective \textit{Aharonov-Bohm} phase. %
% And the appearance of the phase factors $e^{i(\theta _{p,q}-\theta
%_{p+1,q})}$ and $e^{i(\theta _{p,q+1}-\theta _{p,q})}$ in Eq. (2) can be
%understood in terms of the \textit{Aharonov-Bohm} phase accumulated along a
%closed contour \cite{Jean}. So it is important to identify the key element
%for the emergence of artificial gauge potentials: gradient of the phase of
%the external classical driving fields along a single direction (vertical or
%horizontal), and these space-dependent tunneling phase could be
%independently tuned by the local microwave pulses.
With a suitable tuning of the tunneling phases between pairs of neighboring
NVEs, the polariton turns out to experience a non-trivial artificial gauge
potential $\vec{A}$, which can be identified by
\begin{equation}
\theta _{p,q}=\frac{e}{\hbar }\int_{0,q}^{p,q}\vec{A}\cdot d\vec{L}=\frac{e}{%
\hbar }\int_{p,q}^{p,0}\vec{A}\cdot d\vec{L},
\end{equation}%
where the integral is performed along the segment connecting the TLRs, and $%
e $ is the effective charge. With the choice of the symmetric gauge, we set $%
\theta _{p,q}=-\pi pq\alpha $, where $\alpha =\Phi /\Phi _{0}$ with $\Phi $
the magnetic flux through a unit cell, and $\Phi _{0}=h/e$ the flux quantum.
So a uniform artificial gauge field will emerge during the polaritonic
dynamics. The sum of the tunneling phases along a closed loop surrounding
the plaquette is $2(\theta _{p+1,q}+\theta _{p,q+1}-\theta _{p,q}-\theta
_{p+1,q+1})=2\pi \alpha $, which is actually the flux quanta per plaquette,
namely, the strength of the artificial gauge field. Note that the phase
errors of the local microwave source only cause the local gauge field
variance. The total flux remains unchanged.

\begin{figure}[tbp]
\includegraphics[width=5 cm]{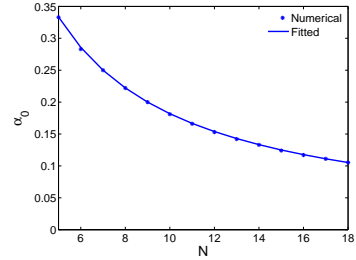}
\caption{(Color online) The relation between the critical value of $\protect%
\alpha _{0}$ and the system size $N$.}
\end{figure}

%In this hybrid system, it is interesting to establish a link between the
%ground states from its polariton distribution among all the lattice sites
%and the so-called Hofstadter butterfly, which describing a fractal structure
%in energy spectrum with respect to magnetic field plane when considering
%non-interacting particles on the periodic lattice moving in a uniform
%magnetic field. In particular, the frustration effect associated with the
%effective gauge field, which is introduced by the position-dependent phase
%factors, can causes the phase of the polariton wave function to be
%nonuniform and induce a degenerate ground state for single-excitatiom case,
%similar to the prediction of Ref. \cite{Moll}.

Next we focus on the observable consequences of a uniform magnetic field in
the present system. For simplicity, we consider the case of non-interacting
polaritons in finite-size $2D$\ square lattice under the tight-binding limit
and the single-polariton subspace. As shown in Fig. 3, we plot the spatial
distribution of polaritons $B_{p,q}$\ in the ground state of the $10\times
10 $\ lattice under different values of $\alpha $. When $\alpha =0,$\ the
Hamiltonian in Eq. (3) can be reduced to that for a free particle in an
infinitely deep square potential well and the form of ground state takes the
Sine functions. When $\alpha $\ increases from $0$\ to $\alpha _{0}\simeq
0.181$, we find that the polaritons concentrate at the central region which
implies the wave-function of the ground state gradually change from Sine
function into that of the $0$-th Landau level with zero angular momentum $m$%
, and the radius of the $0$th Landau level decreases as the gauge field
increases. This transition which is denoted by $\alpha $ is really from 0-th
Landau level of $m=0$ to 0-th Landau level of $m>0$. However the
distribution changes dramatically once the value of $\alpha $ exceeds a
critical value $\alpha _{0}$, where the $0$-th Landau level with $m>0$
momentum becomes ground state. The physical mechanism behind this intriguing
phenomena is the competition between the gauge field and the size effect. As
shown in Fig. 4, we find that the crossing point $\alpha _{0}$\ can be well
fitted by the equation $\alpha _{0}=2/(N+1)$\ for the $N\times N$\ systems
with $N>4$\ (see Sec. III in supplementary material). If we further increase
$\alpha $, many level crossing points would appear. We find that all the
crossing points are near certain rational points $\alpha =p/q,$\ at which
the wave-functions show regular oscillations\cite{HB}. After Fourier
transformation, one can see that such regular oscillations come from the
coherent interference between the peculiar points in momentum space (see
Sec. III in supplementary material). It is very interesting that the results
from the model with the periodic condition in thermodynamic limit can be
observed in this small and open boundary system. In order to measure the
level crossing, we can couple the NVEs to superconducting qubits, and
transfer the state of NVEs to the nearby qubits. Then we can measure the
ground state population distribution of polaritons by measuring the states
of superconducting qubits.

To investigate the possibility of observing the fractal band structure of
HB, we diagonalize the Hamiltonian $H_{B}$ in the single-polariton subspace
for several small lattice sizes. As shown in Fig. 5a, the HB structure can
be observed in the size of $5\times 5$. It becomes clearer in the whole
transmission spectra with the growth of the size. %This intricate
%figure simulates the phase diagram of a charged particle plunged in a
%uniform magnetic field.
Experimentally, transmission and reflection measurements have been by far
performed routinely in small-scale circuit QED systems with one or two
resonators \cite{SC,read}. We can characterize our hybrid circuits by
spectroscopic measurements, and a feasible method is to measure the
resonator transmission through a network analyzer. In realistic experiments,
the transmission spectrum of these superconducting resonators will show a
series of Lorentzian peaks, whose central frequencies correspond to the
eigenfrequencies of the Hamiltonian $H_{eff}$ in Eq. (3). Alternatively,
using homodyne detection followed by sampling and averaging after
amplification, we may reveal the coupled resonator-spin dynamics by
measuring the amplitude of the exponentially damped microwave signal that
leaks out of the resonators after its interaction with the spin ensembles
\cite{Ku}. So far it is still a great challenge to observe such a
fascinating structure in ordinary solid-state system, where the major
challenge comes from the extremely high magnetic field required in realistic
experiments \cite{Cha}. In this sense, our proposal opens an entirely new
arena to investigate condensed matter and many-body system with light.
Meanwhile, the experimental progress, especially in large-scale circuit QED
\cite{Nu1,Nu2}, has raised the possibility of observing the HB spectrum
directly in realistic experiments \cite{expl}.

\begin{figure}[tbph]
\centering \includegraphics[width=8cm,bb=0pt 210pt 594pt
615pt]{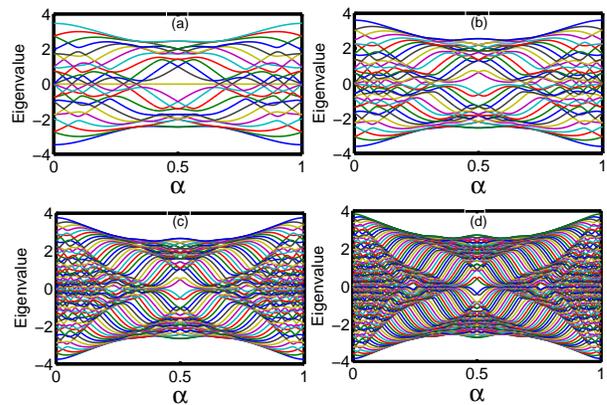}
\caption{(Color online) The energy spectrum for (a) $5\times 5$ lattice; (b)
$6\times 6$ lattice; (c) $8\times 8$ lattice; (d) $10\times 10$ lattice,
where energy is in units of tunneling strength $J$. The vertical axis is the
eigenenergy of the system, and the horizontal axis indicates magnetic flux
through a unit cell. }
\label{fig:2}
\end{figure}

Finally, we survey the relevant experimental parameters. First, the TLR
cavity with an inductance $F_{t}=45.6$\ nH and capacitance $C_{t}=2$\ pF
leads to a full wave frequency $\omega _{0}/2\pi =D_{gs}+\Delta _{c}=3.31$
GHz with $\Delta _{c}/2\pi =440$ MHz. Second, our scheme requires the
large-detuning condition, namely, the detuning $\Delta _{c}\gg g_{eff}$.
Third, the classical field should be tuned to $\Omega /2\pi =400$ MHz $\gg
g_{eff}$, which makes the energy shift induced by the cavity mode negligible
\cite{Fuch}. So if we take the values of the parameters $g_{eff}/2\pi =8$
MHz, $T/2\pi =4$ MHz, and $\delta /2\pi =40$ MHz, yielding $J/2\pi
=Tg_{eff}^{2}/4\delta ^{2}=0.04$ MHz, our scheme is feasible. This could be
confirmed by recent experimental demonstration of coherent coupling of a NVE
and a superconducting flux qubit \cite{XBZhu} as well as the experimental
advances in excellent quantum control with strong magnetic coupling ($\sim
2\pi \times 10$ MHz) between TLR and NVE and the cavity linewidth of $\kappa
/2\pi \sim $ kHz \cite{N1,N2,N3,N4}. On the other hand, the electron spin
relaxation time $T_{1}$ of NV centers ranges from 6 ms at room temperature
\cite{D2} to $28\sim 265$ s at low temperature \cite{D3}. In addition, the
dephasing time $T_{2}$ $>$ 600 $\mu $s for NVE with natural abundance of $%
^{13}C$ has been reported \cite{D4}. A later experimental progress \cite{D5}
with isotopically pure diamond sample has demonstrated a longer dephasing
time to be $T_{2}$ = 1.8 ms. Therefore, the parameter $J$ is higher by
nearly two orders of magnitude than the dissipation rates of the NVE, which
makes reliable quantum simulation feasible.

Compared with previous cavity QED protocols \cite{E2,ZBY}, our scheme
requires only one driving source at each site, and the effective couplings
between neighboring spins are much larger than in Ref. \cite{E2}\ because
the excited states are not required to be adiabatically eliminated in our
case. However, so far we have ignored the detrimental influence from the
nuclear spin, such as $^{13}C$ defects, in the NVE, which could be
alleviated by isotopically purified $^{12}C$ diamond through the
purification technique \cite{N5,D5}. Another decoherence source is the
dipole-interaction between the redundant Nitrogen spins and the NV centers,
which could be reduced by improving the nitrogen to N-V conversion rate
while maintaining the large collective coupling constants \cite{N4}.
Alternatively, this problem could be overcome by applying the external
driving field to the electron spins on the Nitrogen atoms. It would increase
the coherence time of the NVE if these spins are flipped on a time scale
much faster than the flip-flop processes \cite{N5}. On the other hand, the
dephasing time can be greatly enhanced by decoupling the electron spin from
its local environment with a spin echo sequence. Using this technique, the
dephasing time of the NVE reaches 3.7 $\mu$s at room temperature \cite{N4}.

In summary, we have discussed how to simulate gauge field in a J-C lattice
of NVE-TLR system, where the motion of polaritons (collective excitations)
in NVEs is analogous to the motion of charged bosonic particles in a
magnetic field. We have discussed the competition between the size effect
and the gauge field on the ground state crossover, and also discussed the
possibility of identifying signatures of the HB in the optical spectra of
the superconducting resonators. With currently available techniques, we
argue that our system lends itself as a well-suited quantum simulator for
investigating phenomena encountered in condensed matter physics, and our
study would be useful for the future spintronics technology.

We thanks Cong-jun Wu and Hong-hao Tu for valuable discussions. This work is
supported by the NFRP of China under Grants 2011CBA00300, 2011CBA00301, No.
2011CB922102, No. 2011CB921200, No. 2011CBA00200, No. 2011CB921803, No.
2012CB921704, and by NNSF of China under Grants No. 10974225, No. 11004226,
No. 11104326, No. 11105136, No. 11174035, No. 61073174, No. 61033001, and
No. 61061130540, and Chinese PRF under Grant NO. 20110490829, as well as by
the NRFME, Singapore (Grant No. WBS: R-710-000-008-271).


\begin{thebibliography}{99}
\bibitem{expl} Our scheme can also be applied to other spin degrees of
freedom, such as N substitution (P1) centers \cite{N2}, several defects in
silicon carbide (SiC) \cite{SIC}, electronic spins of quantum dot coupled to
microwave superconducting resonator(SC) \cite{QD}, and cold atoms coupled to
SC \cite{Verdu}.
\end{thebibliography}
\end{document}

\title{Supplementary Material}
\author{W. L. Yang$^{1,4}$}
\author{Zhang-qi Yin$^{2,3}$}
\email{yinzhangqi@gmail.com}
\author{Z. X. Chen$^{3}$}
\author{Su-peng Kou$^{5}$}
\author{M. Feng$^{1}$}
\email{mangfeng@wipm.ac.cn}
\author{C. H. Oh$^{4}$}
\email{phyohch@nus.edu.sg}
\affiliation{$^{1}$State Key Laboratory of Magnetic Resonance and Atomic and Molecular
Physics, Wuhan Institute of Physics and Mathematics, Chinese Academy of
Sciences, and Wuhan National Laboratory for Optoelectronics, Wuhan 430071,
China}
\affiliation{$^{2}$Key Laboratory of Quantum Information, University of Science and
Technology of China, Chinese Academy of Sciences, Hefei 230026, China }
\affiliation{$^{3}$Center for Quantum Information, IIIS, Tsinghua University, Beijing,
China}
\affiliation{$^{4}$Centre for Quantum Technologies, National University of Singapore,
Singapore 117543, Singapore}
\affiliation{$^{5}$Department of Physics, Beijing Normal University, Beijing 100875, China}
\maketitle

%\email{ywl@wipm.ac.cn}

\section{Hamiltonian for the single-site NVE-TLR}

We suppose that each NVE couples to a two-mode TLR in each site. By applying
an external resonant driving microwave on the NVE, the Hamiltonian of the $%
(p,q)$-th site reads
\begin{eqnarray}
H_{p,q} &=&\sum_{c=a,b}\omega _{c}c_{p,q}^{\dagger
}c_{p,q}+\sum_{j=1}^{N_{p,q}}\omega _{e}|e\rangle
_{p,q}^{j}{}_{p,q}^{j}\langle e|+[\sum_{j=1}^{N_{p,q}}(\frac{\Omega
_{p,q}e^{i\theta _{p,q}}}{2}e^{-i\omega _{e}t}|e\rangle
_{p,q}^{j}{}_{p,q}^{j}\langle g|  \notag \\
&&+\sum_{c=a,b}g_{p,q}^{c}c_{p,q}|e\rangle_{p,q}^{j}{}_{p,q}^{j}\langle g|)+
\mathrm{H.C.}],
\end{eqnarray}
where $\Omega_{p,q}$ is the Rabi frequency of the driving microwave, and $%
\theta _{p,q}$ is the relative phase. In the rotating frame regarding $%
H_{0}=\sum_{c=a,b}\omega _{e}c_{p,q}^{\dagger }c_{p,q}+\omega
_{e}\sum_{j=1}^{N_{p,q}}|e\rangle _{p,q}^{j}{}_{p,q}^{j}\langle e|$, we get
the Hamiltonian as
\begin{equation*}
H_{p,q}=\sum_{c=a,b}(\omega _{c}-\omega _{e})c_{p,q}^{\dagger
}c_{p,q}+[\sum_{j=1}^{N_{p,q}}(\frac{\Omega _{p,q}e^{i\theta _{p,q}}}{2}
|e\rangle _{p,q}^{j}{}_{p,q}^{j}\langle
g|+\sum_{c=a,b}g_{p,q}^{c}c_{p,q}|e\rangle _{p,q}^{j}{}_{p,q}^{j}\langle
g|)+ \mathrm{H.C.}].
\end{equation*}
Here we define $\Delta _{c}=\omega _{c}-\omega _{e}$ for $c=a,b$.

Under the strong driving case $\Omega _{p,q}\gg g_{p,q}^{c}$, we can define
the new energy levels as $|+\rangle _{p,q}=(e^{i\theta _{p,q}}|e\rangle
_{p,q}+|g\rangle _{p,q})/\sqrt{2}$ and $|-\rangle _{p,q}=(-e^{i\theta
_{p,q}}|e\rangle _{p,q}+|g\rangle _{p,q})/\sqrt{2}$. The Hamiltonian of the
system under the new basis is
\begin{eqnarray}
H_{p,q} &=&\sum_{c=a,b}\Delta _{c}c_{p,q}^{\dagger
}c_{p,q}+\sum_{j=1}^{N_{p,q}}(\frac{\left\vert \Omega _{p,q}\right\vert }{2}
(|+\rangle _{p,q}^{j}{}_{p,q}^{j}\langle +|-|-\rangle
_{p,q}^{j}{}_{p,q}^{j}\langle -|)  \notag \\
&&+\frac{1}{2}\sum_{c=a,b}g_{p,q}^{c}c_{p,q}e^{-i\theta _{p,q}}(|+\rangle
_{p,q}^{j}-|-\rangle _{p,q}^{j})(_{p,q}^{j}\langle +|+_{p,q}^{j}\langle -|)
\notag \\
&&+\frac{1}{2}\sum_{c=a,b}g_{p,q}^{c}c_{p,q}^{\dagger }e^{i\theta
_{p,q}}(|+\rangle _{p,q}^{j}+|-\rangle _{p,q}^{j})(_{p,q}^{j}\langle
+|-_{p,q}^{j}\langle -|).
\end{eqnarray}
Let's neglect the energy shift induced by the cavity mode, which is much
smaller than the energy shift induced by the driving laser. Then we obtain
\begin{eqnarray}
H_{p,q} &=&\sum_{c=a,b}\Delta _{c}c_{p,q}^{\dagger
}c_{p,q}+\sum_{j=1}^{N_{p,q}}(\frac{\left\vert \Omega _{p,q}\right\vert }{2}
(|+\rangle _{p,q}^{j}{}_{p,q}^{j}\langle +|-|-\rangle
_{p,q}^{j}{}_{p,q}^{j}\langle -|)  \notag \\
&&+\frac{1}{2}\sum_{j=1}^{N_{p,q}}\sum_{c=a,b}[g_{p,q}^{c}c_{p,q}e^{-i\theta
_{p,q}}(|+\rangle _{p,q}^{j}{}_{p,q}^{j}\langle -|-|-\rangle
_{p,q}^{j}{}_{p,q}^{j}\langle +|)+\mathrm{H.C.}].
\end{eqnarray}

Suppose $\Delta _{a}\simeq |\Omega _{p,q}|$, $\Delta _{b}\simeq -|\Omega
_{p,q}|$, and $\Delta _{c}$, $\Omega _{p,q}\gg \sqrt{N_{p,q}}g_{p,q}^{c}$,
we can obtain the effective Hamiltonian under the rotating wave
approximation,
\begin{eqnarray}
H_{p,q}^{eff} &=&\delta _{p,q}^{a}a_{p,q}^{\dagger }a_{p,q}+\delta
_{p,q}^{b}b_{p,q}^{\dagger }b_{p,q}+\frac{1}{2}
\sum_{j=1}^{N_{p,q}}[g_{p,q}^{a}a_{p,q}e^{-i\theta _{p,q}}|+\rangle
_{p,q}^{j}{}_{p,q}^{j}\langle -|  \notag \\
&&-g_{p,q}^{b}b_{p,q}^{\dagger }e^{-i\theta _{p,q}}|+\rangle
_{p,q}^{j}{}_{p,q}^{j}\langle -|+\mathrm{H.C.}],
\end{eqnarray}
where $\delta _{p,q}^{a}=\Delta _{a}-\Omega _{p,q}$ and $\delta _{p,q}^{b}
=\Delta _{b}+\Omega _{p,q}$. Let's define $\tilde{S}_{p,q}^{+}=%
\sum_{j=1}^{N_{p,q}}|+\rangle _{p,q}^{j}\langle -|$ and $\tilde{S}%
_{p,q}^{-}=\sum_{j=1}^{N_{p,q}}|-\rangle _{p,q}^{j}\langle +|$ for $(p,q)$%
-th site. We denote the Hamiltonian of the coupling between mode $a$ and NVE
by $H_{p,q}^{a}=\delta _{p,q}^{a}a_{p,q}^{\dagger }a_{p,q}+( \frac{1}{2}%
g_{p,q}^{a}e^{-i\theta _{p,q}}a_{p,q}\tilde{S}_{p,q}^{+}+\mathrm{\ H.C.})$.
On the other hand, the effective Hamiltonian describing the coupling between
mode $b$ and NVE is given by $H_{p,q}^{b}=\delta _{p,q}^{b}b_{p,q}^{\dagger
}b_{p,q}-(\frac{1}{2}g_{p,q}^{b}e^{-i\theta _{p,q}}b_{p,q}^{\dagger }\tilde{S%
}_{p,q}^{+}+\mathrm{H.C.}).$

In the rotating frame regarding $H_{0}=\delta _{p,q}^{a}a_{p,q}^{\dagger
}a_{p,q}+\delta _{p,q}^{b}b_{p,q}^{\dagger }b_{p,q}$, we rewrite Eq.(4) as
\begin{equation}
H_{p,q}^{eff}=\frac{1}{2}g_{p,q}^{a}e^{-i(\theta _{p,q}+\delta
_{p,q}^{a}t)}a_{p,q}\tilde{S}_{p,q}^{+}-\frac{1}{2}g_{p,q}^{b}e^{-i(\theta
_{p,q}-\delta _{p,q}^{b}t)}b_{p,q}\tilde{S}_{p,q}^{+}+\mathrm{H.C.}.
\end{equation}

\section{Effective Hamiltonian for coupled cavity sites}

The coupling between $(p,q)$-th site and the nearest TLR sites can be
modeled by following Hamiltonian.
\begin{equation}
H_{T}=-\sum_{p,q=1}^{N}\hbar (T_{a}a_{p,q+1}^{\dagger
}a_{p,q}+T_{a}a_{p,q}^{\dagger }a_{p,q+1}+T_{b}b_{p+1,q}b_{p,q}^{\dagger
}+T_{b}b_{p,q}b_{p+1,q}^{\dagger }).  \label{eq:Htran}
\end{equation}
We diagonalize the Hamiltonian \eqref{eq:Htran} with the Fourier
transformation $a_{k_{q},p}=\frac{1}{\sqrt{N}}\sum_{q=1}^{N}e^{-i2\pi
k_{q}q/N}a_{p,q}$ and $b_{k_{p},q}=\frac{1}{\sqrt{N}}\sum_{p=1}^{N}e^{-i2\pi
k_{p}p/N}b_{p,q}$,
\begin{equation}
H_{T}^{\prime }=-\sum_{k_{p}=2\pi /N}^{2\pi
}\sum_{q=1}^{N}T_{k_{p}}b_{k_{p},q}^{\dagger }b_{k_{p},q}-\sum_{k_{p}=2\pi
/N}^{2\pi }\sum_{p=1}^{N}T_{k_{p}}a_{k_{p},p}^{\dagger }a_{k_{p},p},
\label{eq:Htran1}
\end{equation}
where $T_{k_{p}}=2T_{b}\cos (2\pi k_{p}/N)$ and $T_{k_{q}}=2T_{a}\cos (2\pi
k_{q}/N)$. We define $F_{p,k_{p}}=e^{-i2\pi k_{p}p/N}$ and $%
F_{q,k_{q}}=e^{-i2\pi k_{q}q/N}$.

It is found that the coupling between vertical sites and between horizonal
sites can be calculated independently. The effective Hamiltonian of the
system for the vertical line $q$ is
\begin{equation}
H_{v}=-\sum_{p=1}^{N}(\frac{1}{2}g_{p,q}^{b}e^{-i(\theta _{p,q}-\delta
_{p,q}^{b}t)}b_{p,q}\tilde{S}_{p,q}^{-}+\mathrm{H.C.})-\sum_{k_{p}=2\pi
/N}^{2\pi }T_{k_{p}}b_{k_{p},q}^{\dagger }b_{k_{p},q}.  \label{eq:Hv}
\end{equation}%
For simplicity, we assume $\delta _{p,q}^{a}=\delta _{p,q}^{b}=\delta $, $%
g_{p,q}^{a}=g_{p,q}^{b}=g$, and $T_{p,q}=T$. With the transformation $%
b_{k_{p},q}=\sum_{p=1}^{N}F_{k_{p},p}b_{p,q}$, we get
\begin{equation*}
H_{v}=-\sum_{p=1}^{N}\sum_{k_{p}=2\pi /N}^{2\pi }(\frac{1}{2\sqrt{N}}%
gF_{k_{p},p}b_{k_{p},q}\tilde{S}_{p,q}^{-}e^{-i\theta _{p,q}}e^{i(T+\delta
)t}+\mathrm{H.C.}).
\end{equation*}%
In the limit of $\delta \gg T,g$, we can adiabatically eliminate the cavity
modes $b_{k_{p},q}$ and get
\begin{equation}
H_{v}^{eff}=\sum_{p,p^{\prime }}^{N}\frac{g^{2}}{4}e^{i(\theta _{p^{\prime
},q}-\theta _{p,q})}\tilde{S}_{p^{\prime },q}^{+}\tilde{S}%
_{p,q}^{-}\sum_{k_{p}=2\pi /N}^{2\pi }\frac{F_{k_{p}^{\prime
}}F_{k_{p}}^{\ast }}{N}\frac{1}{\delta +T_{k_{p}}}.  \label{eq:Hv1}
\end{equation}%
In the limit of $\delta \gg T$, we get $1/(\delta +T_{k_{p}})\simeq \frac{1}{%
\delta }-\frac{T_{k_{p}}}{\delta ^{2}}$, $%
\sum_{k_{p}}F_{k_{p},p}F_{k_{p},p^{\prime }}^{\ast }=N\delta _{p,p^{\prime
}} $, and $\sum_{k_{p}}F_{k_{p},p}F_{k_{p},p^{\prime }}^{\ast
}T_{k_{p}}=T(\delta _{p,p^{\prime }+1}+\delta _{p+1,p^{\prime }})$.
Neglecting the Stark shift terms, we get the effective Hamiltonian as
\begin{equation}
H_{v}^{eff}=\sum_{p,p^{\prime }}^{N}\frac{Tg^{2}}{4\delta ^{2}}e^{i(\theta
_{p+1,q}-\theta _{p,q})}\tilde{S}_{p+1,q}^{+}\tilde{S}_{p,q}^{-}+\mathrm{H.C.%
}  \label{eq:Hv2}
\end{equation}

Similarly, we can calculate the effective Hamiltonian for the horizontal
line $p$ by adiabatically eliminating the cavity modes as the following
form,
\begin{equation}
H_{h}^{eff}=\sum_{q,q^{\prime }}^{N}\frac{Tg^{2}}{4\delta ^{2}}e^{i(\theta
_{p,q}-\theta _{p,q+1})}\tilde{S}_{p,q+1}^{+}\tilde{S}_{p,q}^{-}+\mathrm{H.C.%
}  \label{eq:Hh}
\end{equation}%
Combining Eqs. \eqref{eq:Hv} and \eqref{eq:Hh}, we obtain the effective
Hamiltonian for the whole system as
\begin{equation}
\begin{aligned} H_{\mathrm{eff}}=&-J^{'}(\sum_{p,q}
e^{i(\theta_{p,q}-\theta_{p+1,q})}\tilde{S}_{p,q}^+ \tilde{S}_{p+1,q}^- \\&+
e^{i(\theta_{p,q+1}-\theta_{p,q})}\tilde{S}_{p,q}^+ \tilde{S}_{p,q+1}^- +
\mathrm{H.C.}), \end{aligned}
\end{equation}%
where $J^{^{\prime }}=T(g/2\delta )^{2}$.

\section{The ground state near level crossing}

\begin{figure}[tbph]
\centering\includegraphics[width=12cm]{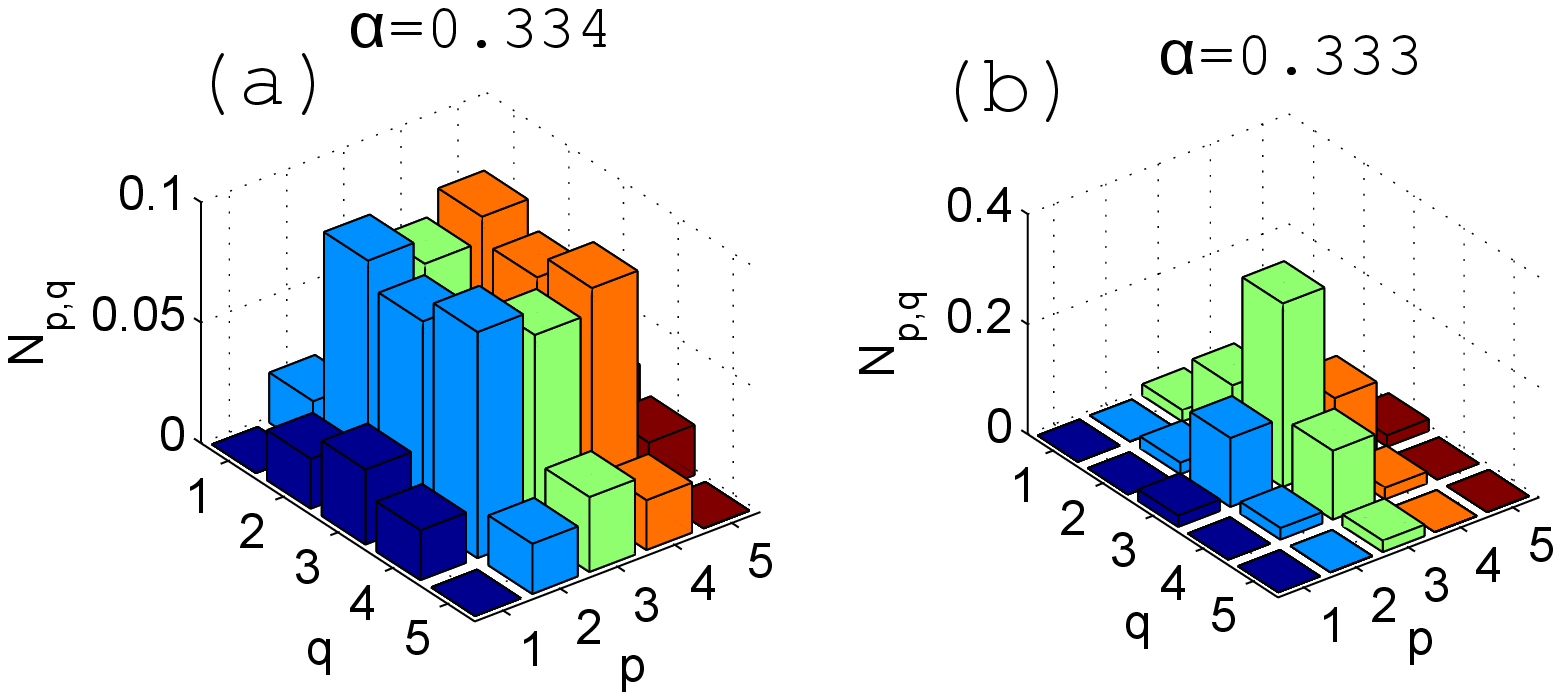}
\caption{(Color online) The spatial distribution of the NVE polaritons for a
$5\times 5$ lattice under different values of $\protect\alpha$. (a) $\protect%
\alpha =0.334$. (b) $\protect\alpha =0.333$. }
\label{fig:s1}
\end{figure}

We find that the ground state level crossing can be observed in the system
as small as $5\times 5$ lattices. As shown in Fig.\ref{fig:s1}, the critical
value of gauge field is $\alpha_0=0.333\simeq 1/3$.

The numerical result of the ground state near the critical gauge field $%
\alpha _{0}$ for the $5\times 5$ lattice system is shown as follows,
\begin{equation}
\begin{aligned} \Phi(0.333) =&
0|1\rangle_{1,1}+(0.0003-0.0002i)|1\rangle_{1,2}+(0.144-0.0009i)|1%
\rangle_{1,3} \\&+(0.0003+0.0002i)|1\rangle_{1,4}+0|1\rangle_{1,5} \\
&+(0.0003+0.0002i)|1\rangle_{2,1}+0.1448|1\rangle_{2,2}+(0.3536-0.0011i)|1%
\rangle_{2,3}
\\&+(0.1448-0.0009i)|1\rangle_{2,4}+(0.0003-0.0002i)|1\rangle_{2,5}\\
&+(0.1440+0.0009i)|1\rangle_{3,1}+(0.3536+0.0011i)|1\rangle_{3,2}+0.5772|1%
\rangle_{3,3}
\\&+(0.3536-0.0011i)|1\rangle_{3,4}+(0.144+0.0009i)|1\rangle_{3,5}\\
&+(0.0003-0.0002i)|1\rangle_{4,1}+(0.1448+0.0009i)|1\rangle_{4,2}%
\\&+(0.3536+0.0011i)
|1\rangle_{4,3}+0.1448|1\rangle_{4,4}+(0.0003+0.0002i)|1\rangle_{4,5}\\
&+0|1\rangle_{5,1}+(0.0003+0.0002i)|1\rangle_{5,2}+(0.1440+0.0009i)|1%
\rangle_{5,3} \\&+(0.0003-0.0002i)|1\rangle_{5,4}+0|1\rangle_{5,5}.
\end{aligned}  \label{111}
\end{equation}%
\begin{equation}
\begin{aligned} \Phi(0.334) =&
-0.0004|1\rangle_{1,1}+(0.1251-0.00713i)|1\rangle_{1,2}+(0.0022-0.1763i)|1%
\rangle_{1,3} \\&+(-0.1232+0.00745i)|1\rangle_{1,4}+0.0004|1\rangle_{1,5} \\
&+(0.1251+0.0713i)|1\rangle_{2,1}+0.3065|1\rangle_{2,2}+(0.0018-0.2890i)|1%
\rangle_{2,3}
\\&+(0.3064-0.0039i)|1\rangle_{2,4}+(0.1260-0.0698i)|1\rangle_{2,5}\\
&+(0.0022+0.1763i)|1\rangle_{3,1}+(0.0018+0.2890i)|1\rangle_{3,2}+0|1%
\rangle_{3,3}
\\&+(-0.0018+0.2890i)|1\rangle_{3,4}+(-0.0022+0.1763i)|1\rangle_{3,5}\\
&+(-0.1232+0.0745i)|1\rangle_{4,1}+(-0.3064+0.0039i)|1\rangle_{4,2}%
\\&+(-0.0018-0.2890i)
|1\rangle_{4,3}+0.3065|1\rangle_{4,4}+(0.1242+0.00729i)|1\rangle_{4,5}\\
&+0.0004|1\rangle_{5,1}+(-0.1260+0.0698i)|1\rangle_{5,2}+(-0.0022+0.1763i)|1%
\rangle_{5,3} \\&+(0.1242-0.0729i)|1\rangle_{5,4}-0.0004|1\rangle_{5,5}.
\end{aligned}  \label{222}
\end{equation}%
Here $|\Psi (\alpha )\rangle $ denotes the ground state at the field $\alpha
$, and $|1\rangle _{p,q}$ denotes the Fock state $|1\rangle $ of the
polariton at the lattice site $p,q$ . It is easy to verify that $\langle
\Phi (0.333)|\Phi (0.334)\rangle =1.58\times 10^{-5}\sim 0$. Therefore,
there is level crossing near the critical gauge field $\alpha _{0}=0.333$.

In general, we can define the ground state fidelity for the system $F(\alpha
)=\langle \Phi (\alpha )|\Phi (\alpha +\delta )\rangle $, with $\delta
\rightarrow 0$. For those $F(\alpha )$ approach zero, the level crossing
occurs. We have numerically calculated the $F(\alpha )$ and identified the
level crossing points for different sizes of lattices.
%\emph{In Fig. 2, we have plotted the }$F(\alpha )$\emph{\ for
%different size of lattices.
For example, for $5\times 5$ lattice, the level crossing point is near $1/3$%
. For $6\times 6$ lattice, the level crossing points are near $2/7$, $3/8$, $%
2/5$ and $1/2$. The critical points are fitted very well with rational
number $p/q$ for different lattices, and the first crossing point $\alpha
_{0}$\ is just $2/(L+1)$\ for the $L\times L$\ systems ($L>4$). To
qualitatively learn the nature of the first level crossing at $\alpha
_{0}=2/(L+1),$ we estimate $\alpha _{0}$ by perturbation approach.

First we consider a zero magnetic field case, $\alpha =0$. Now the effective
model in Eq.(12) is reduced to a continuum form as
\begin{equation}
H_{B}\rightarrow \frac{\hat{k}_{p}^{2}}{2m}+\frac{\hat{k}_{q}^{2}}{2m},
\end{equation}%
where $m=\frac{1}{2J}$. Considering the boundary condition,
\begin{align*}
\Psi (0,0)& =0,\text{ }\Psi (L_{p},0)=0, \\
\Psi (0,L_{q})& =0,\text{ }\Psi (L_{p},L_{q})=0,
\end{align*}%
where $L_{p}$ ($L_{q}$) is the number sites along p(q) direction. We have
the wave-function as
\begin{equation*}
\Psi _{(p,q)}=\frac{2}{\sqrt{L_{p}L_{q}}}\sin (\frac{\pi p}{L_{q}})\sin (%
\frac{\pi q}{L_{p}}),
\end{equation*}%
and the energy eigenvalues as
\begin{equation*}
E_{(p,q)}=\frac{1}{2m}[(\frac{\pi p}{L_{p}})^{2}+(\frac{\pi q}{L_{q}})^{2}],
\end{equation*}%
where $p=1,2,3,..L_{p}$ and $q=1,2,3,..L_{q}$. It means that the ground
state for the $\alpha \rightarrow 0$ case is labeled by $p=1$, $q=1$ with
the energy
\begin{equation*}
E_{(1,1)}=\frac{1}{2m}[(\frac{\pi p}{L_{p}})^{2}+(\frac{\pi q}{L_{q}})^{2}].
\end{equation*}%
The first excited state is labeled by $p=2,$ $q=1$ or $p=1,$ $q=2$ with the
energy
\begin{equation*}
\min (E_{(2,1)},E_{(1,2)})=\min (\frac{1}{2m}[(\frac{2\pi p}{L_{p}})^{2}+(%
\frac{\pi q}{L_{q}})^{2}],\frac{1}{2m}[(\frac{\pi p}{L_{p}})^{2}+(\frac{2\pi
q}{L_{q}})^{2}]).
\end{equation*}

Secondly we add a tiny magnetic field $\alpha $. The energy spectrum of the
system changes little. So we may use the perturbation theory to deal with
this case. Here we choose the Landau gauge, $\vec{A}=(By,0)$, where the
gauge field strength $B=2\pi \alpha$. In the following part we set $%
L_{x}=L_{x}=L$.

Now the model in Eq.(12) in a continuum form turns into
\begin{equation}
H_{B}^{\prime }\rightarrow \frac{(\hat{k}_{p}-Bq)^{2}}{2m}+\frac{\hat{k}%
_{q}^{2}}{2m}.
\end{equation}%
In the perturbation theory, we use the plane waves of non-magnetic case $%
\Psi _{(p,q)}$ to describe above model. Then for the lowest eigenstate $\Psi
_{(1,1)},$ and the excited state $\Psi _{(2,1)},$ we have the corresponding
perturbation energies as
\begin{equation*}
E_{(1,1)}^{\prime }=\frac{1}{2m}[(\frac{\pi }{L})^{2}]+\frac{1}{2m}[(\frac{%
\pi }{L})^{2}]+\frac{1}{L^{2}}\frac{1}{2m}\sum_{q=1}^{L}2\frac{\pi }{L}%
(-2\pi \alpha \ q)+\alpha ^{2}\ \text{term,}
\end{equation*}%
and
\begin{equation*}
E_{(2,1)}^{\prime }=\frac{1}{2m}[(\frac{\pi }{L})^{2}]+\frac{1}{2m}[(\frac{%
2\pi }{L})^{2}]+\frac{1}{L^{2}}\frac{1}{2m}\sum_{q=1}^{L}2\frac{2\pi }{L}%
(-2\pi \alpha \ q)+\alpha ^{2}\ \text{term,}
\end{equation*}%
where the $\alpha ^{2}\ $terms can be neglected when $\alpha $ $\ll 1$. So
there may exist level crossing between $\Psi _{(1,1)}$ and $\Psi _{(2,1)}$
when
\begin{equation*}
E_{(1,1)}^{\prime }=E_{(2,1)}^{\prime }
\end{equation*}%
that is
\begin{equation*}
\frac{1}{2m}[(\frac{2\pi }{L})^{2}]-\frac{1}{2m}[(\frac{\pi }{L})^{2}]=\frac{%
1}{L^{2}}\frac{1}{2m}\sum_{q=1}^{L}2\frac{2\pi }{L}(2\pi \alpha _{c}\ q)-%
\frac{1}{L^{2}}\frac{1}{2m}\sum_{q=1}^{L}2\frac{\pi }{L}(2\pi \alpha _{c}\ q)
\end{equation*}%
for tiny $\alpha .$ That is
\begin{equation*}
\frac{3}{2m}(\frac{\pi }{L})^{2}=\frac{1}{L^{2}}\frac{1}{2m}\sum_{q=1}^{L}2%
\frac{\pi }{L}(2\pi \alpha _{c}\ q)
\end{equation*}%
or%
\begin{equation*}
\alpha _{c}=\frac{3}{2}\frac{1}{(L+1)}
\end{equation*}%
with $\sum_{q=1}^{L}q=\frac{L(L+1)}{2}.$ Therefore we estimate the first
level crossing at $\alpha _{0}=1.5/(L+1)$ by perturbation approach, which is
qualitatively same as the numerical result $\alpha _{0}=2/(L+1)$.

\section{Interference\textbf{\ pattern for }$\protect\alpha =p/q$}

\begin{figure}[tbp]
\includegraphics[width=0.7\textwidth]{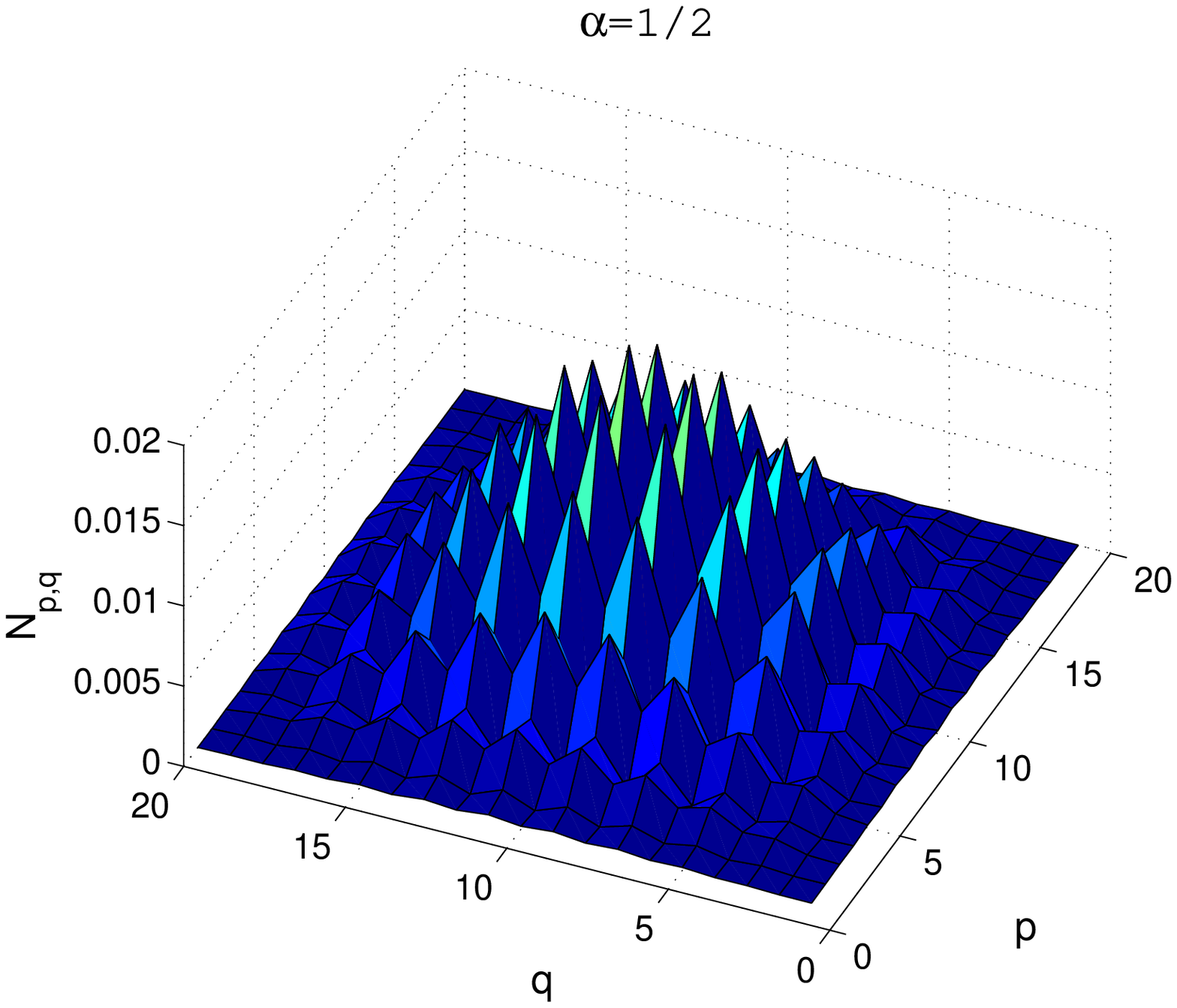}
\caption{The total fermion density of $\protect\pi $-flux case on $20\times
20$ lattice. }
\end{figure}

From the ground state fidelity we have found that all the level crossing
points are near certain rational points $\alpha =p/q,$ at which the
wave-functions show regular oscillations. Fig. 2 illustrates the particle
density of $\alpha =\frac{1}{2}$ case on $20\times 20$ lattice. After
Fourier transformation, we find that such regular oscillations come from the
coherent interference between the peculiar points in momentum space. Thus
for the ground state of $\alpha =\frac{1}{2}$ case on an $L\times L$ lattice
with open boundary condition, the particle density can be transformed into $%
8 $ resonant points inside the reduced Brillouin zone (BZ) of momentum space
as
\begin{equation*}
(k_{p},k_{q})=(\frac{\pi }{2}\pm \frac{\pi }{L},\pm \frac{\pi }{2}\pm \frac{%
\pi }{L})
\end{equation*}
around the nodal points $(k_{p},k_{q})=(\frac{\pi }{2},\pm \frac{\pi }{2})$.
Let's explain this result. For an infinite system, the case of $\alpha =%
\frac{1}{2}$ is just a $\pi $-flux model, of which there are two nodal
points in the reduced BZ as $(\frac{\pi }{2},\pm \frac{\pi }{2}).$ To
transform the wave-function of the system with finite size, we can consider
the $L\times L$ lattice with open boundary condition as an infinite system
with periodic $L\times L$ super-cells, of which the wave-functions are fixed
to be zero at the interfaces between two $L\times L$ super-cells. Thus,
modulated by $L\times L$ super-cells, each nodal points split to $4$ points.
As a result,\ we can read out the information of a model with the periodic
condition in thermodynamic limit in this small and open boundary system. As
shown in Fig. 3, for a $20\times 20$ lattice, the $8$ resonant points in the
reduced BZ zone are already very close to the $2$ nodal points for an
infinite system.

\begin{figure}[tbp]
\includegraphics[width=0.5\textwidth]{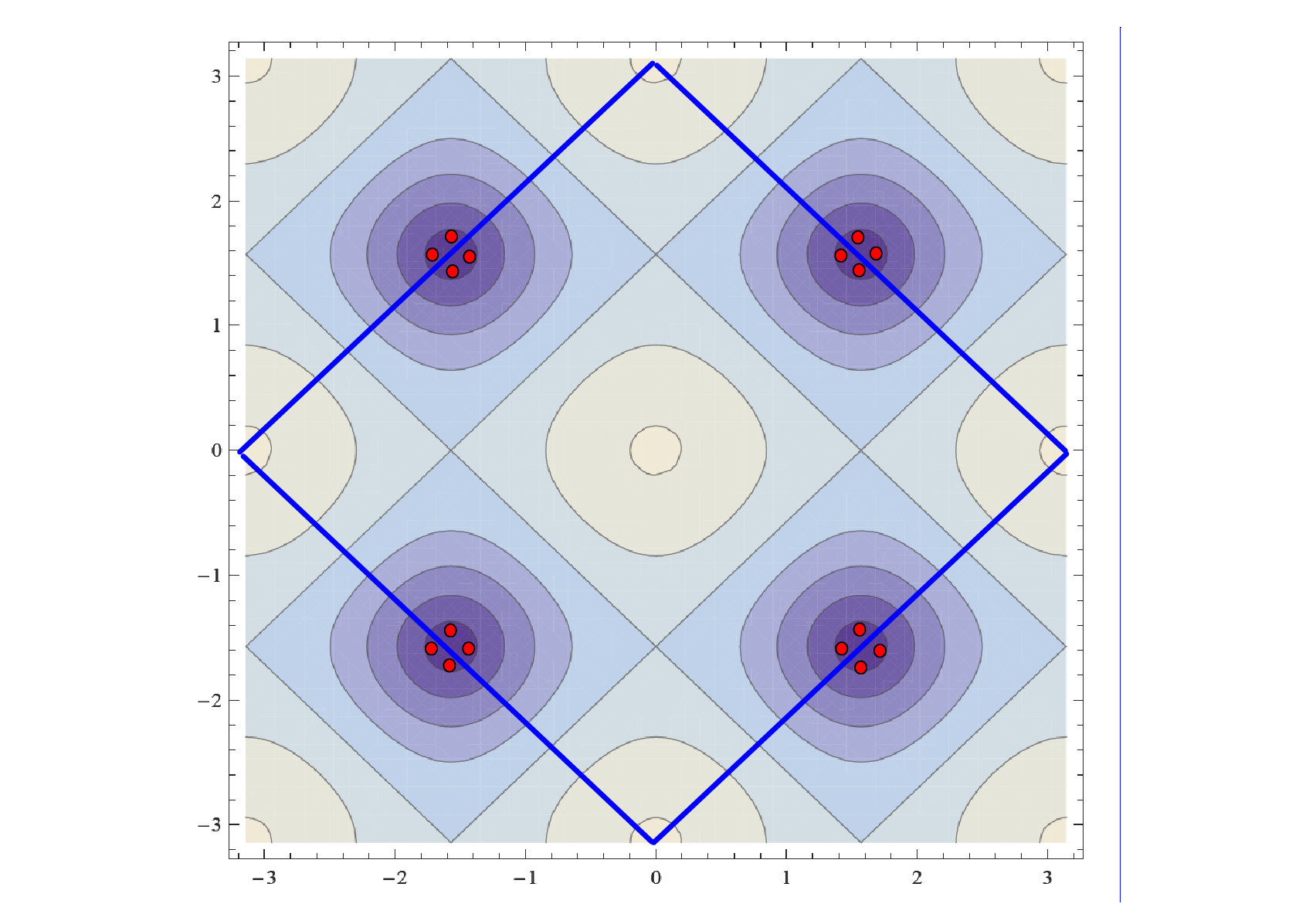}
\caption{The illustration of interference pattern of $\protect\pi $-flux
case on $20\times 20$ lattice in momentum space. The blue lines are the
boundary of reduced BZ. The 8 resonant points is $(k_{p},k_{q})=(\frac{%
\protect\pi }{2}\pm \frac{\protect\pi }{20},\pm \frac{\protect\pi }{2}\pm
\frac{\protect\pi }{20}).$}
\end{figure}